\pdfoutput=1

\documentclass[11pt]{article}

\usepackage[]{ACL2023}

\usepackage{times}
\usepackage{latexsym}

\usepackage[T1]{fontenc}

\usepackage[utf8]{inputenc}

\usepackage{microtype}

\usepackage{inconsolata}
\usepackage{multirow}
\usepackage{tabularx}
\usepackage{booktabs}
\usepackage[most]{tcolorbox}
\usepackage{pifont}
\usepackage{amssymb}

\usepackage{afterpage}
\title{Detecting Calls to Action in Multimodal Content: Analysis of the 2021 German Federal Election Campaign on Instagram}

\author{Michael Achmann-Denkler \\
Media Informatics Group \\ University of Regensburg \\ Regensburg, Germany \\
  \texttt{michael.achmann@ur.de} \\\And
  Jakob Fehle \\
Media Informatics Group \\ University of Regensburg \\ Regensburg, Germany \\
  \texttt{jakob.fehle@ur.de} \\\AND
  Mario Haim \\
Department of Media and Communication \\ Ludwig-Maximilians-Universität \\ Munich, Germany \\
  \texttt{haim@ifkw.lmu.de} \\\And
    Christian Wolff \\
Media Informatics Group \\ University of Regensburg \\ Regensburg, Germany \\
  \texttt{christian.wolff@ur.de} \\
  }

\begin{document}
\maketitle
\begin{abstract}
This study investigates the automated classification of Calls to Action (CTAs) within the 2021 German Instagram election campaign to advance the understanding of mobilization in social media contexts. We analyzed over 2,208 Instagram stories and 712 posts using fine-tuned BERT models and OpenAI’s GPT-4 models. The fine-tuned BERT model incorporating synthetic training data achieved a macro F1 score of 0.93, demonstrating a robust classification performance. Our analysis revealed that 49.58\% of Instagram posts and 10.64\% of stories contained CTAs, highlighting significant differences in mobilization strategies between these content types. Additionally, we found that FDP and the Greens had the highest prevalence of CTAs in posts, whereas CDU and CSU led in story CTAs.
\end{abstract}

\section{Introduction}


In this study, we experiment with the automated classification of Calls to Action (CTAs) from the 2021 German Instagram campaign to advance the understanding of mobilization in social media election campaigns. Our primary goal is to determine the efficacy of several computational approaches for binary classification of the presence or absence of CTAs in Instagram posts and stories from the 2021 Federal election in Germany. To this end, we fine-tuned a BERT model \cite{devlin-etal-2019-bert}, experimented with synthetic training data to enhance the model, and contrasted these approaches with zero- and few-shot prompting using OpenAI's GPT-4 model family. Through our study, we aim to address the three gaps in computational text analysis for the social sciences identified by \citet{Baden2022-oe}: 1) We experiment with a non-English language, 2) We evaluate all classifications against human annotations for external validation \cite{Birkenmaier2023-nt}, and 3) We investigate the potential of LLMs for overcoming the specialization before integration gap.

The 2021 election marked a shift in Germany’s political landscape, with the long-serving Chancellor Angela Merkel stepping down. The key parties in the race included the CDU/CSU, SPD, Greens, FDP, AfD, and The Left. In 2021, Instagram was used by almost the same share of the German population as Facebook and was particularly popular among younger users under the age of 30 \cite{Koch2022}. About half of the candidates had profiles on Instagram, with notable differences between parties \cite{Kelm2023-uu}. We are interested in the front-runner and party accounts and how they utilized CTAs on Instagram to gain insight into their mobilization and audience engagement strategies. Understanding these strategies reveals how political actors use Instagram to engage voters. Thus, our secondary goal is to use the CTA classifications to contrast mobilization strategies between Instagram stories and posts, filling a gap as ephemeral stories have often been overlooked. Therefore, we want to answer the following research questions:

\begin{description}
\item[RQ1a] \emph{Which of the currently available GPT-4 model variants, when tested with few-shot and zero-shot prompts, achieves the highest performance in automated detection of CTAs in German-language Instagram content?}
\item[RQ1b] \emph{Does incorporating synthetic training data enhance the performance of a fine-tuned BERT model in detecting CTAs in German-language Instagram content?}
\item[RQ1c] \emph{When comparing the best-performing GPT and BERT models, what are the performance differences in detecting CTAs between different types of Instagram content (stories vs. posts) and text types (OCR vs. caption vs. transcript)?}
\item[RQ2] \emph{How does the usage of CTAs vary between different types of Instagram content (stories vs. posts) and between different political parties?}
\end{description}

\subsection{Political Communication on Instagram}
Instagram's role in political communication has been extensively studied, addressing various political actors and nations. Studies commonly reveal that political figures use Instagram to project positive imagery rather than for policy discussion or voter engagement \cite{Bast2021-fh}. Studies of the 2021 German Federal election have focused on visual personalization and political issues in posts \cite{Schlosser2023-tu, Hasler2023-kr, Geise2024-un}, and Instagram stories were compared to regular posts using topic modeling \cite{Achmann2023-rs}. 

Voter engagement and mobilization on social media have been the focus of recent studies: \citeauthor{Magin2017-uz} (\citeyear{Magin2017-uz}) illustrated that about half of the posts in the 2013 German and Austrian election campaigns on Facebook included CTAs, primarily focusing on mobilization. \citeauthor{Larsson2024-is} (\citeyear{Larsson2024-is}) proposed a framework for comparing political actors' campaign strategies across social media platforms. They investigated the Norwegian parliamentary election campaign on three social media platforms: Facebook, Instagram, and Twitter. \citeauthor{Wurst2023-pn} (\citeyear{Wurst2023-pn}) examined the mobilization strategies used by German political parties during the 2021 election campaign on Facebook and Instagram. Their findings revealed that 43\% of Instagram posts from parties and candidates included mobilization calls. The study found notable differences in mobilization strategies among parties, with the Greens using calls to vote more frequently than others.




The current research offers a comprehensive view of how CTAs are used in social media campaigns. This paper aims to extend the analysis to include both Instagram posts and Stories, offering a more holistic view of political campaigning on this platform.

\subsection{Ephemeral Instagram Stories}
Few studies have investigated ephemeral Instagram stories in the context of political campaigns and communication: \citet{Towner2022-oe} analyzed stories from 2020 U.S. presidential candidates. They collected 304 images one week before and after the election campaign. They found the campaigns missed opportunities to share user-generated content and inconsistently followed communication norms for Instagram Stories. \citeauthor{Towner2024-gr} (\citeyear{Towner2024-gr}) studied how gubernatorial candidates utilized Instagram Stories during the 2018 elections. They found that candidates primarily used stories to mobilize voters and showcase indoor events, preferring static images to videos. This area remains relatively unexplored compared to the analysis of Instagram posts.



\subsection{Text-Mining in Political Communication}
Textual analysis of Instagram content includes a frequency study to analyze Islamist extremist content \cite{Clever2023-fc}, and an analysis of political advertisements on Instagram and Facebook, utilizing computational text classification methods \cite{Vargo2020-rs}.

The computational detection of CTAs in social media content has, for example, been investigated by \citeauthor{Rogers2019-ik} (\citeyear{Rogers2019-ik}). They classified CTAs on VKontakte, focusing on their role in mobilization and potential for censorship. Their model demonstrates a classification performance of F1=0.77. They used a relatively small ground-truth dataset (n=871) and employed RuBERT, a Russian version of BERT. Similarly, \citeauthor{Siskou2022-md} (\citeyear{Siskou2022-md}) developed a rule-based Natural Language Processing (NLP) pipeline to identify CTAs in Spanish social media posts. Their approach yields F1 scores between 0.81 and 0.85. \citeauthor{Gupta2020PoliBERT} (\citeyear{Gupta2020PoliBERT}) report in their working paper on training a fine-tuned BERT model for classifying political tweets and Facebook posts from the 2016 US General Election. They achieved an F1 score of 0.92 for CTAs on Twitter and 0.95 for Facebook. 

In conclusion, these studies highlight the potential of using advanced NLP approaches and BERT variants to detect political CTAs in different languages and social media platforms.

\subsection{Large Language Models for Social Science Tasks}
LLMs have shown proficiency in various text classification tasks, including social sciences tasks, with some studies indicating performance superior to human annotators \cite{Liu2023-bi, Tornberg2023-lu, Gilardi2023-uu}.
While they are promising for tasks with clear and well-defined criteria, such as identifying misinformation or distinguishing political stances, applying LLMs requires caution, particularly in tasks needing deep semantic understanding \cite{Ziems2023-tk}.


Beyond prompting, LLMs may also be used to augment training data: \citeauthor{Bertaglia2024-gx} (\citeyear{Bertaglia2024-gx}) explored using GPT-3.5 Turbo to generate synthetic Instagram captions for detecting sponsored content. Combining synthetic with real data improved their classification F1 score from 0.71 to 0.78, demonstrating that synthetic data can enhance classifier training.

In summary, Instagram is a critical platform for political communication. Prior research validates the potential of advanced NLP models, including BERT variants and LLMs, for detecting CTAs. Our study aims to compare GPT-4 and a fine-tuned BERT model to classify CTAs in German Instagram texts, using synthetic training data for enhanced performance. 

\section{The Corpus}
We collected two types of Instagram content: permanent posts that may include multiple images or videos with a caption and stories that typically consist of a single image or video. Captions in posts represent the primary textual content on Instagram, varying in length and often featuring hashtags. While captions are the primary text elements, many images and videos incorporate embedded text or spoken words.

For our computational analysis, we deconstructed each Instagram post and story into smaller units to analyze text in various forms: captions, embedded text (through Optical Character Recognition, OCR), and speech (transcriptions) for video audio. This approach resulted in up to two text documents per image and up to three documents per video. As a post can contain multiple images, this leads to a maximum of $3 \cdot n_{\text{images}}$ documents per post, plus an additional document for the caption. In contrast, Instagram stories typically comprise a single image or video, resulting in one OCR document and an optional transcription document per story. See table \ref{tab:corpus} for an overview of corpus statistics for each text type, and table \ref{tab:examples} for examples.

\begin{table}[]
\caption{Corpus statistics grouped by post- and text-type.}
\label{tab:corpus}
\resizebox{\linewidth}{!}{%
\begin{tabular}{@{}ll | cc | ccc@{}}
\toprule
\textbf{Post Type} & \textbf{Text Type}  & \multicolumn{2}{ c |}{\textbf{Documents}} &  \multicolumn{3}{ c }{\textbf{Tokens}} \\ 
& & \textbf{\#} & \textbf{\%} & \textbf{\#} & \textbf{mean} & \%  \\ \midrule
Post       & Caption        & 720          & 15.57        & 48449     & 67.29       & 34.02     \\ 
Post       & OCR            & 1093         & 23.64        & 15529     & 14.21       & 10.90      \\
Post       & Transcription  & 138          & 2.99         & 22099     & 160.14      & 15.52     \\ \midrule
Story      & OCR            & 2157         & 46.66        & 41850     & 19.40        & 29.38     \\
Story      & Transcription  & 515          & 11.14        & 14499     & 28.15       & 10.18     \\ \midrule
\multicolumn{2}{r |}{\textbf{Overall}} & 4623         & 100          & 142426    & 289.19      & 100      \\ \bottomrule
\end{tabular} }%
\end{table}

\begin{table*}
\caption{A sample of text documents and their human annotations for the presence (\ding{52}) or absence (\ding{56}) of Call to Action (CTA).}
\label{tab:examples}
\small
\begin{tabularx}{\linewidth}{l l | X | c | c}
\toprule
\textbf{Post Type}  & \textbf{Text Type} & \textbf{Example} & \textbf{CTA} & \textbf{Username} \\ \midrule

Post & Caption & Jede*r vierte Erwerbstätige arbeitet für weniger als 12 Euro pro Stunde. Das reicht selbst bei Vollzeitarbeit kaum zum Leben. Deshalb sorgen wir für höhere Löhne und gesunde Arbeitsbedingungen. Denn Arbeit muss gerecht bezahlt werden. Du willst, dass alle Menschen von ihrer Arbeit leben können. Dann wähl Grün am Sonntag. & \ding{52} & @die\_gruenen \\ \midrule

Post & OCR & ROT-ROT-GRUN WURDE FUR MILLIONEN MENSCHEN IN BAYERN EINE VERSCHLECHTERUNG DER LEBENSSITUATION BEDEUTEN. MARKUS SÖDlR CSU & \ding{56} & @markus.soeder \\ \midrule


Story & Transcription & Nicht verpassen, heute einschalten, einundzwanzig Uhr fünfzehn, Home Sweet Germany mit mir. & \ding{52}  & @cdu \\ \bottomrule

\end{tabularx}
\end{table*}

\subsection{Data Collection \& Preprocessing}
We collected stories and posts published by eight parties, namely \textit{AfD} (@afd\_bund), \textit{CDU} (@cdu), \textit{CSU} (@christlichsozialeunion), \textit{Die Grünen} (@die\_gruenen), \textit{Die Linke} (@dielinke), \textit{FDP} (@fdp), \textit{FW} (@fw\_bayern), and \textit{SPD} (@spdde), and 14 front-runners\footnote{We only collected stories from verified accounts. In case of missing accounts or verification marks, we followed the hierarchy Chancellor-Candidate $>$ Front-Runner $>$ Head of Party $>$ Deputy Head of Party. CDU and CSU are running a joint campaign; therefore, just one candidate each is included.} (see table~\ref{tab:politicians} in the appendix). Data collection started two weeks before election day, from Sept. 12th until Sept. 25, 2021, excluding election day. During this time, parties and politicians shared 712 posts and 2208 stories. Posts were collected retrospectively using CrowdTangle, amounting to 1153 images and 151 videos. Stories were collected daily at 0:00 using the \texttt{selenium} Python package to simulate a human user browsing the stories.\footnote{We can not guarantee completeness for Sep 14 due to technical problems.} A majority of the posted stories are videos (n=1246). 

Many images contain embedded text, which we extracted using OCR (\texttt{easyocr}). We transcribed videos using the \textit{whisper-large-v2-cv11-german} model,\footnote{\href{https://huggingface.co/bofenghuang/whisper-large-v2-cv11-german}{https://huggingface.co/bofenghuang/whisper-large-v2-cv11-german}} a version of OpenAI's Whisper model \citep{Lucas2022-cs} fine-tuned for German. We also applied OCR to the first frame of videos.


\section{Methods}
We have operationalized CTAs as a binary variable, indicating their presence or absence in documents, simplifying our model's classification process. Each social media post or story is analyzed by decomposing it into several text documents, enabling the computational analysis of multimodal data. To answer questions on a post/story level, we assign `True' for an entire post or story if Call to Action is marked as `True' in any of the associated documents. This section defines CTAs, describes our annotation study, and the prompt engineering and model training steps.

\subsection{Calls to Action}
A ``Call to Action'' (CTA) refers to statements or prompts that explicitly encourage the audience to take immediate action \cite{Ilany_Tzur2016-bd}. \citeauthor{Larsson2024-is} (\citeyear{Larsson2024-is}) connect CTAs in political campaigns to three of \citeauthor{Magin2017-uz}'s (\citeyear{Magin2017-uz}) campaign functions: Informing, Mobilizing, and Interacting. The first function aims at disseminating messages and positions on important issues. Mobilizing encourages supporters to take active steps such as voting, participating in events, or sharing campaign messages. Interacting facilitates dialogue between politicians and citizens, enhancing engagement and potentially persuading voters more effectively through reciprocal communication \cite{Magin2017-uz}. \citeauthor{Wurst2023-pn} (\citeyear{Wurst2023-pn}) relate to these functions and define three types of CTA: ``Calls to Inform'' encourage the audience to seek further online or offline information. This could include directing users to the party's website or inviting them to read party-related materials. ``Calls to Interact'' aim to increase engagement through dialogue, such as inviting users to comment on a post or participate in discussions. Finally, ``Calls to Support'' are direct appeals for actions that benefit the party, such as voting, donating, or sharing posts to increase the campaign's visibility.

We consider CTAs as a dichotomous variable marking the presence or absence of \textit{any} CTA in a document. While this reduction from three types into a singular CTA reduces the analytical value of our work, we see it as a simplification to create a robust classification model. Such a model can then be used to develop more nuanced classification models in future studies.


\subsection{The Annotation Process}
Preparing our corpus, we drew a stratified sample across text (caption, OCR, transcript) and content type (story, post) combinations. The documents were annotated across two batches: We started with a 20 \% sample in the first batch (n=925) and increased the sample size to 1,388 documents (app. 30 \%) through a second batch.\footnote{Overall, our text corpus comprises 4,614 documents; sample sizes were rounded when balancing the text- and content-type distribution.} Each document was independently annotated by at least three randomly assigned annotators. A total of nine annotators contributed to the annotation. Alongside one of the authors who participated in the annotation process, we recruited eight non-expert annotators from our staff and students. The latter were rewarded with participant hours for their work. The majority (8) of annotators were native German speakers. Participants received a detailed annotation guide, including examples and the GPT classification prompt (see appendix, figure \ref{fig:prompt-cta}). They had to pass a short quiz to ensure they read the manual before being invited to the annotation project. Annotations were collected remotely using the Label Studio software. Participants coded one document at a time, marking the presence of CTAs with ``True'' or ``False''. ``Unsure'' responses were coded as \texttt{NA}.

Items with disagreement were passed into a second round of annotations to increase the number of votes. Overall, nine coders created 5290 annotations. Using a majority decision, we deduced the ground truth CTA labels. Ties were resolved through the author's annotation. The interrater agreement measured by Krippendorff's $\alpha$ reached a moderate level of $\alpha$=0.67 \cite{Krippendorff2004-mt}. Notably, the agreement between the majority decisions and the annotating author reached a strong level \cite{McHugh2012-ua}, with Cohen's $\kappa$=0.88 (n=892, excluding ties) \cite{Cohen1960-ot}. This alignment with the author’s labels confirms the validity of our final dataset, demonstrating that the majority decision effectively captures \textit{Calls to Action}, despite the expected variability among non-expert student annotators.


\begin{table}[]
\caption{An overview of the annotated corpus. About one-fifth of text documents contain (\ding{52}) Calls to Action.}
\resizebox{\linewidth}{!}{%
\begin{tabular}{@{}ll | cc | c @{}}
\toprule
\textbf{Post Type}  & \textbf{Text Type}      & \multicolumn{2}{c |}{\textbf{\ding{52}}}   & \textbf{\ding{56}} \\ \midrule
Post       & Caption        & 106 & (49.30\%) & 109    \\
Post       & OCR            & 52  & (15.85\%) & 276    \\
Post       & Transcription  & 11  & (26.19\%) & 31     \\ \midrule
Story      & OCR            & 91  & (14.04\%) & 557    \\
Story      & Transcription  & 8   & (5.16\%)  & 147    \\ \midrule
\multicolumn{2}{r |}{\textbf{Overall}} & 268 & (19.31\%) & 1,120   \\ \bottomrule
\end{tabular} }%
\end{table}

\subsection{Classification Approaches}
We compare several classification approaches using transformer architectures and large language models to detect the presence of \textit{Call for Actions} within posts and stories shared during the election campaign. Specifically, we compare two main classification methods: fine-tuning the gbert-large German BERT model and utilizing OpenAI's GPT-4 large language model. We tested different variations for each method: we trained two BERT models—one with the original dataset and another with an extended dataset augmented by GPT-4o. For the GPT approach, we tested GPT-4, GPT-4 Turbo, and GPT-4o models in both zero-shot and few-shot settings.


\subsection{Fine-tuned BERT models}
We fine-tuned the pre-trained `deepset/gbert-large` model for our German language classification task using the \texttt{tansformers} library \cite{wolf_2022_7391177}. GBERT is a state-of-the-art BERT model trained on German text \cite{chan-etal-2020-germans}. We trained two classification models: \textbf{gbert-cta} trained on the original dataset, and \textbf{gbert-w/-synth-cta} trained on the original dataset + synthetic data generated using GPT-4o to mitigate the class imbalance of the original dataset.

Both models went through the same preprocessing and training steps. Input documents were tokenized, with truncation and padding to a maximum length of 512 tokens. The training took place on Google Colab, using Nvidia A100 graphics cards. We used wandb\footnote{\href{https://wandb.ai/site}{wandb.ai}} to find the best hyperparameters, focusing on achieving the highest F1 score. To address the class imbalance in the gbert-cta model, we calculated class weights and added them to the loss function. After optimizing the hyperparameters, we validated each model with a five-fold cross-validation. This means we split the dataset into five parts stratified by the call to action variable, trained the model in four parts, and tested it on the remaining part. We added one-fifth of the synthetic data to the training data per fold for the model incorporating the synthetic dataset. We repeated this process five times, each with a different part as the test set, ensuring a robust evaluation.

\subsection{Synthetic Dataset}
To improve the quality of our BERT classification model, we generated synthetic data to counter the class imbalance of our ground truth dataset. We generated three synthetic texts for each of the documents classified to contain a CTA using the prompt in the appendix, see figure \ref{fig:prompt-synth}. During the training of the gbert-w/-synth-cta model\footnote{Available at \href{https://huggingface.co/chaichy/gbert-CTA-w-synth}{https://huggingface.co/chaichy/gbert-CTA-w-synth}}, we appended the synthetic data to the training set, paying attention to not leaking any synthetic data into the evaluation dataset and, vice-versa, to not leak any evaluation or test data through synthetic data based on these datasets, into the training data. We used the following parameters for our API requests: \texttt{gpt-4o-2024-05-13},
 \texttt{temperature=0} and \texttt{top\_p=1}. The \texttt{max\_tokens} were set individually: We calculated the number of tokens for each original text using the \texttt{tiktoken} package provided by OpenAI and used the original token count as \texttt{max\_tokens}.

\subsection{Zero- and Few-Shot using GPT}
Following \citeauthor{Tornberg2024-ub}'s (\citeyear{Tornberg2024-ub}) recommendations, we initiated the prompt engineering process by having one author annotate a small random sample of 150 documents. Next, we hand-crafted a preliminary classification prompt: ``Given any user input, classify whether the input contains any calls to action''. We tested the initial draft on ChatGPT to classify one document at a time. Responding to misclassifications, we provided nuanced examples and instructed ChatGPT to modify and improve the original prompt accordingly.\footnote{for example: \href{https://chatgpt.com/share/fdd306b0-ff2d-4971-bad2-92eb6e8f07a7}{https://chatgpt.com/share/fdd306b0-ff2d-4971-bad2-92eb6e8f07a7}} Thus, we started to improve the prompt by conversation with GPT-4.0 on the ChatGPT platform. Once the classifications on ChatGPT appeared satisfactory, we used the prompt with the API and inferred classifications for all 150 sampled captions. This iterative prompt development process has been previously demonstrated to be effective \cite{Pryzant2023-ch}. Through the iterations, we added examples, as few-shot prompts have also been proven effective \cite{Brown2020-il}. 

During this prompt optimization process, we compared the classification results to the author's annotations and calculated Cohen's $\kappa$ as a benchmark for the prompt's quality. Ultimately, we settled on a prompt incorporating \citeauthor{Tornberg2024-ub}'s advice to construct prompts around context, the question, and constraints. The context was provided in the objective part of the prompt, the question in the instructions part, and the constraints in the formatting part. Additionally, we enumerated the instructions and potential types of CTAs. Within the instructions, we employed the chain-of-thought approach \cite{Wei2022-op}, as the model was prompted to split input messages into sentences, classify each sentence, and then return the final classification. See figure \ref{fig:prompt-cta} in the appendix for the final result. We deleted the examples from the few-shot prompt to convert it into the zero-shot prompt.

Our commands were sent as system prompts to the API, while each document was sent as user messages. We used the following settings for our API requests: 
 \texttt{temperature=0}, \texttt{max\_tokens=5}, and \texttt{top\_p=1}. We used the following model versions: \texttt{gpt-4-0613}, \texttt{gpt-4-turbo-2024-04-09}, and \texttt{gpt-4o-2024-05-13}.

\subsection{Evaluation Approach}
We evaluated our classification approaches using established machine learning evaluation metrics: precision, recall, macro F1-score, and binary F1-score. The metrics were calculated using \texttt{scikit-learn} \cite{scikit-learn}. Additionally, we calculated Cohen's $\kappa$ to measure the interrater agreement between our ground truth data and the model classifications for comparison with social science research.

We used an independent test dataset to evaluate our BERT model. The corpus was stratified by "Call to Action" and split into two sets: 80\% for training and 20\% for testing. The 80\% training set was used for hyperparameter tuning and cross-validation, while the 20\% test set was reserved for the final evaluation. To evaluate the GPT classifications, we excluded rows containing phrases from the few-shot examples (n=16) and used the entire annotated dataset.

\section{Results}
In the first part of this section, we will answer our primary questions \textbf{RQ1a--c} regarding the computational classifications through the external evaluation based on human annotations. At the end of the section, we will answer our secondary interest \textbf{RQ2}, uncovering the differences between stories, posts, and parties.

\subsection{Evaluation of GPT Models}
\begin{table}
\centering
\caption{Evaluation of CTA detection across different GPT-4 model variations and prompt types (few-shot vs. zero-shot). The highest values are marked in bold.}
\label{tab:results}
\resizebox{\linewidth}{!}{%
\begin{tabular}{@{}lcc|cc|cc@{}}
\toprule
 & & &  \multicolumn{2}{c|}{\textbf{$F_1$}} &  &  \\ 

Model       & Prompt      & $\kappa$  & Macro       & Binary      & Precision      & Recall         \\ \midrule
GPT-4o      & Few  & \textbf{0.81} & \textbf{0.91} & \textbf{0.84} & 0.82          & 0.87          \\
GPT-4 Turbo & Few  & 0.77          & 0.89          & 0.81          & 0.72          & \textbf{0.92} \\
GPT-4       & Few  & 0.80          & 0.90          & 0.84          & \textbf{0.95} & 0.75         \\ \midrule
GPT-4o      & Zero & 0.79          & 0.90          & 0.82          & 0.86          & 0.78          \\
GPT-4 Turbo & Zero & \textbf{0.81} & \textbf{0.90} & \textbf{0.84} & 0.85          & \textbf{0.83} \\
GPT4        & Zero & 0.70          & 0.85          & 0.76          & \textbf{0.94} & 0.64          \\ \bottomrule
\end{tabular} }%
\end{table}
The performance across all tested GPT models is consistently high: The macro F1 scores\footnote{Subsequently, we always refer to macro F1 scores unless stated otherwise.} range from F1=0.85 to F1=0.91 (compare table \ref{tab:results}). GPT-4o, with the few-shot prompt, achieves the highest classification performance, answering \textbf{RQ1a}. Upon closer inspection, the model performs best when classifying captions, followed by OCR in posts and post transcriptions. For stories, the performance drops to F1=0.85 for OCR and even lower for transcription text. 



\subsection{Evaluation of BERT Models}
Both BERT models display a comparatively high classification quality ranging from F1=0.92 for the model trained on the original data to F1=0.93 for the model incorporating the synthetic training data (see table \ref{tab:bert-models}). Thus, to answer \textbf{RQ1b}: incorporating synthetic training data generated by GPT-4o improved classification performance. Since the performance has only improved by the second decimal place, the synthetic text generation prompt should be revisited to introduce greater linguistic variety, and the overall results should be interpreted with caution. The small quality improvement might be influenced by other factors, suggesting that the answer to RQ1b is not universally valid. A five-fold cross-validation evaluated the model hyperparameters. The mean F1=0.90 score for the gbert-w/-synth-cta model demonstrates its ability to generalize well across different subsets of the data, and the standard deviation of 0.02 suggests a stable performance with minimal variability. 



\begin{table}[]
\caption{Classification metrics on the independent test dataset for the fine-tuned gbert models.}
\label{tab:bert-models}
\resizebox{\linewidth}{!}{%
\begin{tabular}{@{}l|cc|cc@{}}
\toprule
 & \multicolumn{2}{c|}{\textbf{$F_1$}} &  &  \\ 
\textbf{Model Name} & \textbf{Macro} & \textbf{Binary} & \textbf{Precision} & \textbf{Recall} \\ \midrule
\textbf{gbert-cta}          & 0.92 & 0.87 & 0.86 & 0.89    \\ 
\textbf{gbert-w/-synth-cta} & \textbf{0.93} &\textbf{ 0.89} & \textbf{0.98} & \textbf{0.81} \\ \bottomrule
\end{tabular}
}%
\end{table}

\subsection{Performance Across Text-Type and Post-Type Combinations}
To answer \textbf{RQ1c}, we investigated the classification performance for each text-type and post-type combination (compare table \ref{tab:details}). Notably, the poor results for the classification of story transcriptions and the excellent results for post transcriptions stand out. These outliers may be partly attributed to the low number of cases: of the 12 post transcriptions in the test set, one contains a call to action. Both models classified the document correctly; the F1 score is perfect without false positives. However, across story transcriptions, the BERT model missed three out of four CTAs across 26 documents. Coincidentally, two out of the three false negatives are Calls to Interact. They have been neglected in posts of the 2021 campaign \cite{Wurst2023-pn}, indicating that the training data contains few documents of this type.

%

\begin{table*}[]
\centering
\caption{Evaluation results per document and post type combination for the best classification models for the test dataset. The highest values between models are marked in bold.}
\label{tab:details}
\begin{tabular}{@{}ll|ccc|ccc|c@{}}
\toprule
          &               & \multicolumn{3}{c|}{\textbf{gbert-w/-synth}} & \multicolumn{3}{c|}{\textbf{GPT-4o}} &  \\
Post Type & Text Type     & $\kappa$ & $F_1$ Macro &  $F_1$  Binary & $\kappa$ & $F_1$ Macro &  $F_1$  Binary & n   \\ \midrule
Post      & OCR           & \textbf{0.81} & \textbf{0.91} & \textbf{0.83}  & 0.73 & 0.87 & 0.75  & 59  \\
Post      & Caption       & \textbf{0.86} & \textbf{0.93} & \textbf{0.92}  & \textbf{0.86} & \textbf{0.93} & \textbf{0.92}  & 44  \\
Post      & Transcription & \textbf{1}     &  \textbf{1}       &  \textbf{1}        &  \textbf{1}       &  \textbf{1}       &  \textbf{1}       & 12  \\ \midrule
Story     & OCR           & \textbf{0.91} & \textbf{0.95} &\textbf{ 0.92}  & 0.78 & 0.89 & 0.81  & 137 \\
Story     & Transcription & 0.36 & 0.67 & 0.4    & \textbf{0.76} & \textbf{0.88} & \textbf{0.80}  & 26  \\ \bottomrule
\end{tabular} 
\end{table*}

The lower classification performance of GPT-4o across OCR texts compared to the BERT model is striking. Across both post types, OCR documents constitute about 70\% of all text documents and show the lowest mean token count per document. The OCR process introduces noise by recognizing irrelevant text, i.e., street and shop signs in the background and incorrectly recognized words. The OCR text bits are concatenated and do not necessarily follow the right word order. For captions, the OpenAI model is on par with the BERT model and exceeds the fine-tuned model in transcriptions.  

\subsection{Calls to Action in Posts and Stories}


We used the gbert-w/-synth-cta classifications to answer \textbf{RQ2}: Instagram posts display a higher relative mention of Calls to Action. Almost half of all captions (44.7\%) contain CTAs, followed by 16.8\% of transcriptions and 15.9\% of OCR documents. In stories, we found the most CTAs in the embedded text (10.5\%) and a very low number in transcriptions (2.3\%). On the post/story level, almost half of all posts contain a Call to Action (49.58\%), compared to only 10.64\% of all stories. The difference between CTAs in posts and stories is significant ($\chi^2$(1) = 501.84, p < .001), with a medium effect size (Cramer's V = 0.42).

Next, we tested the use of CTAs across parties for all post types: The analysis indicates a significant difference in their usage between different parties ($\chi^2$(15) = 604.13, p < .001), with a medium effect size (Cramer’s V = 0.46). We accounted for the interaction between party and post type to ensure this difference was not due to varying distributions of post types between parties. This suggests that the parties varied in their use of calls to action in their Instagram election campaigns, even considering the different use of stories and posts across parties. For posts, the FDP displayed the highest use of CTAs (70.45\%), followed by the Greens (60.23\%). On the low end, the SPD made the least use of CTAs (31.97\%), followed by AfD (40.54\%). In stories, the parties acted differently: The CDU (18.76\%) and CSU (14.78\%) show the highest use of calls. Similarly, the Freie Wähler party (14.97\%) and the Left (14.56\%) display relatively high numbers of CTAs in their stories. The CTA leaders for posts, the Greens (5.12\%) and the FDP (5.66\%), are at the bottom of the list for stories. 


\section{Discussion}
Our experiments confirm the efficacy of large language models for the binary classification of Calls to Action in social media election campaigns. Overall, the GPT-4 models performed well in zero- and few-shot settings. Regarding Cohen's Kappa, there is a strong agreement between language model classifications and ground truth labels. 

Fine-tuning the gbert-large BERT model, however, exceeds the performance of the LLMs. The relatively low number of 1,388 human-annotated documents, with 270 positive cases, yielded a well-performing classification model. Adding synthetic training data generated by the GPT-4o model improved the model further. Both models surpass the performance of CTA classification approaches reported for Russian \cite{Rogers2019-ik} and Spanish \cite{Siskou2022-md} social media texts. Compared to a fine-tuned version of BERT for classifying CTAs in English Twitter and Facebook messages \cite{Gupta2020PoliBERT}, our models perform similarly well while using only a third of the training data. 

A closer look at the classification quality on a text- and post-type level reveals problems with classifying story transcripts. CTAs in these documents account for only 5.16\% of the overall training data. This highlights the potential for further improvements in data augmentation using synthetic documents: A qualitative inspection of synthetic training data generated based on transcripts revealed less similarity to original transcripts than to, for example, post captions. Improving the synthetic data prompt to generate more realistic transcripts might improve the classification performance for this type of text while increasing the linguistic variance across synthetic training data might further increase the overall classification performance. 

Striving for the best possible annotation quality, we chose the gbert-w/-synth-cta for our classification task. However, training a robust classification model takes several steps, from annotation through hyperparameter tuning to the final evaluation. Conversely, the GPT-4 models are readily available, and prompt engineering was comparatively uncomplicated in our context. With decreasing prices, evolving models, and the availability of open-source alternatives, like Llama 3, this study further confirms the utility of large language models for computational social science tasks and political science analyses. 

After applying the model, we uncovered significant differences between political actors' use of CTAs in stories and posts. We found a slightly higher prevalence of CTAs across posts compared to previous studies \cite{Wurst2023-pn}, which may be attributed to our sample: We collected data close to election day, CTAs have been shown to increase closer to election day \cite{Stromer-Galley2021-yk, Wurst2023-pn}. Our study contributes to the study of election campaigns mainly by uncovering a significant difference between posts and stories and between parties. The Greens, for example, have been highlighted before as the party with the highest prevalence of CTAs across their posts. At the same time, we found the party's stories contain the lowest number of CTAs relative to the number of stories posted. Overall, the use of CTAs in stories was low, which contrasts with \citeauthor{Towner2024-gr}'s (\citeyear{Towner2024-gr}) observations of the 2018 U.S. gubernatorial election, raising questions about what other elements or content constituted political stories in the 2021 election. 



\subsection{Limitations}
Our study has several limitations. We observed the campaign for a relatively short period -- two weeks -- due to the necessary effort to capture ephemeral stories. Additionally, we limited our study to verified accounts only. We also limited the analysis to the first frame of each video to decrease complexity, possibly dismissing embedded text in any other frame. 


\subsection{Future Work}
The literature on calls to action in election campaigns distinguishes between different types of CTAs that fulfill various campaign functions. To gain a more holistic understanding of election campaigns and increase the analytical power of our approach, we see future work to build on top of our classification model: Using the positive classifications, future studies can collect human annotations to train a multi-label classification model. Following \citeauthor{Tornberg2024-ub}'s argumentation, future work should evaluate the classification performance of open-source LLMs.


\subsection{Ethical Considerations}
We collected publicly available data posted by parties and verified party officials only. We followed the recommendations towards a conscientious approach to data collection by Venturini and Rogers, who considered scraping a ``necessary evil'' \cite{Venturini2019-ve}. In our article, we do not address personal or sensitive data. 

\bibliography{custom}
\bibliographystyle{acl_natbib}

\appendix

\section{Appendix}
\label{sec:appendix}
$\rightarrow$ See next page.

\begin{table*}
\centering
\caption{Selected politicians' accounts and their positions and party affiliation at the time of data collection.} 
\label{tab:politicians}

\begin{tabular}{@{}llll@{}}
\toprule
Name                    & Party     & Position             & Username                   \\ \midrule
Alice Weidel            & AfD       & Front-Runner         & @alice.weidel             \\
Jörg Meuthen            & AfD       & Head of Party        & @joerg.meuthen            \\
Armin Laschet           & CDU       & Chancellor Candidate & @armin\_laschet           \\
Markus Söder            & CSU       & Head of Party        & @markus.soeder            \\
Annalena Baerbock       & GRÜNE     & Chancellor Candidate & @abaerbock                \\
Robert Habeck           & GRÜNE     & Front-Runner         & @robert.habeck            \\
Ates Gürpinar           & Die Linke & Deputy Head of Party & @atesgurpinar             \\
Susanne Henning-Wellsow & Die Linke & Head of Party        & @susanne\_hennig\_wellsow \\
Christian Lindner       & FDP       & Front-Runner         & @christianlindner         \\
Nicola Beer             & FDP       & Deputy Head of Party & @nicola\_beer             \\
Engin Eroglu            & FW        & Deputy Head of Party & @engin\_eroglu            \\
Gregor Voht             & FW        & Deputy Head of Party & @grey\_gor                \\
Olaf Scholz             & SPD       & Chancellor Candidate & @olafscholz               \\
Saskia Esken            & SPD       & Head of Party        & @saskiaesken              \\ \bottomrule
\end{tabular}
\end{table*}

\begin{figure*}
    \begin{tcolorbox}[enhanced, drop shadow, colback=white, sharp corners=southwest, rounded corners=northeast, boxrule=1pt, width=\textwidth]
    \begin{minipage}{\textwidth}
    \footnotesize
Review the text below. It is a \{text\_type\} \{post\_type\} from the 2021 German Federal Election campaign, shared by one of the political parties. Human annotators identified calls to action in this text, which may be explicit or implicit. These calls to action could, for example, include urging viewers to vote for a particular party, attend an event, or visit a website for more information. \\

Your task is to generate an additional text that mimics the style, type, and features of the provided example. The text will be used as synthetic examples to train a BERT model, so it must be representative and diverse. \\

\#\# Task Details: \\
- The text should clearly fit the defined content type: \{post\_type\}. \\
- The style should align with the descriptor: \{text\_type\}. \\
- The example has been posted by the Instagram user \{row['username']\}, representative of the party \{row['party']\} \\
- The length of the generated text should match the length of the example below (approx. \{len(example)\} characters). \\
- Text text should incorporate exactly one call to action in each text \\

\#\# Instructions: \\
1. Analyze the provided example to maintain consistency in tone and style, and party affiliation. \\
2. Include one distinct call to action in each generated text. \\
3. Tailor each text to the context of the 2021 German Federal Election campaign. \\
4. Produce all texts in German to maintain authenticity and relevance to the election context. \\
5. Ensure that each text aligns with the political affiliation with \{row['party']\} to maintain variety and minimize bias in the training dataset. \\

\#\# Formatting: \\
- Output should consist solely of the generated \{post\_type\} \{text\_type\} texts. \\
- Do not include any additional text, commentary, or formatting elements in your response. \\

\#\# Example: \\
\{example\}

    \end{minipage}
    \end{tcolorbox}
      \caption{The text generation prompt used with GPT-4o to generate synthetic training data.}
  \label{fig:prompt-synth}
\end{figure*}

\begin{figure*}
    \begin{tcolorbox}[enhanced, drop shadow, colback=white, sharp corners=southwest, rounded corners=northeast, boxrule=1pt, width=\textwidth]
    \begin{minipage}{\textwidth}
    \footnotesize
    You're an expert in detecting calls-to-action (CTAs) from texts. \\
    
    \#\#Objective: \\
    Determine the presence or absence of explicit and implicit CTAs within German-language content sourced from Instagram texts such as posts, stories, video transcriptions, and captions related to political campaigns from any user input. \\
    
    \#\#Instructions: \\
    1. Examine each input message. \\
    2. Segment the content into individual sentences. \\
    3. For each sentence, identify: \\
    \hspace*{0.5cm} a. Explicit CTA: Direct requests for an audience to act which are directed at the reader, e.g., "beide Stimmen CDU!", "Am 26. September \#FREIEWÄHLER in den \#Bundestag wählen." \\
    \hspace*{0.5cm} b. Explicit CTA: A clear direction on where or how to find additional information, e.g., "Mehr dazu findet ihr im Wahlprogramm auf fdp.de/vielzutun", "Besuche unsere Website für weitere Details." \\
    \hspace*{0.5cm} c. Implicit CTA: Suggestions or encouragements that subtly propose an action directed at the reader without a direct command, e.g., "findet ihr unter dem Link in unserer Story." \\
    4. CTAs should be actions that the reader or voter can perform directly, like voting for a party, clicking a link, checking more information, etc. General statements, assertions, or suggestions not directed at the reader should not be classified as CTAs. \\
    5. If any CTA is detected return 'True'. Otherwise, return 'False'. \\
    
    \#\#Formatting: \\
    Just return your classification result, either True or False.
    \end{minipage}
    \end{tcolorbox}
      \caption{The few-shot CTA detection prompt. It was converted into the zero-shot prompt by deleting the examples.}
  \label{fig:prompt-cta}
\end{figure*}

\end{document}